\documentclass[10pt]{iopart}

\usepackage{iopams}
\expandafter\let\csname equation*\endcsname\relax 
\expandafter\let\csname endequation*\endcsname\relax
\usepackage{amsmath}
\usepackage{xcolor} 
\usepackage{url,hyperref,booktabs,natbib, multirow}
\usepackage[utf8]{inputenc}
\usepackage{newunicodechar}
\newunicodechar{₂}{$_2$}
\usepackage[onehalfspacing]{setspace}
\usepackage{graphicx}
\usepackage{algorithm, algpseudocode}
\usepackage{mdframed}
\usepackage{array}

\usepackage{tabularx}

\newif\ifshowrev
\showrevtrue

\begin{document}

\title{Two-State Dynamical Synapses as a Physical Substrate for Equilibrium Propagation}

\author{Dongshu Liu\,$^{1,*}$, Julie Grollier\,$^{1,*}$}

\address{$^{1}$Laboratoire Albert Fert, CNRS, Thales, Université Paris-Saclay, Palaiseau, France}

\ead{dongshu.liu@cnrs.fr, julie.grollier@cnrs-thales.fr}
\vspace{10pt}
\begin{indented}
\item[]December 2025
\end{indented}


\begin{abstract}

Equilibrium Propagation (EqProp) updates each synapse by comparing a local term at the free and nudged equilibria. Because these two values occur at different times, a physical implementation requires temporary local memory. We show that this memory can emerge from a two-timescale synaptic model in which neural and volatile synaptic dynamics are jointly derived from a single energy function. A volatile state $w$ relaxes toward the product of the pre- and postsynaptic activities, while a non-volatile state $M$ stores the learned parameter. At equilibrium, $w$ equals the local term entering the EqProp update. Using its net change between phases implements phase-based EqProp, while transferring its successive changes during nudging yields continual EqProp, in which the temporal comparison is carried by the synaptic dynamics themselves. On MNIST with a single-hidden-layer network, the phase-based model reaches $2.01 \pm 0.09\%$ test error versus $2.00 \pm 0.13\%$ for standard EqProp, and the continual model reaches $2.24 \pm 0.11\%$ versus $2.28 \pm 0.16\%$ for continual EqProp. The framework further specifies the dynamical requirements that a volatile--non-volatile synaptic device must satisfy to implement this learning mechanism.

\end{abstract}

\vspace{2pc}
\noindent{\it Keywords}: Equilibrium propagation, Dynamic synapses, Neuromorphic Computing, Volatile memory, Non-volatile memory,  Local learning rule

\section{Introduction}
Local learning rules, in which a synapse is modified only by the activity of adjacent neurons, are attractive for neuromorphic hardware because they avoid moving parameters between a memory array and an external processor \citep{Querlioz2013, srinivasan2016magnetic, hansen2018unsupervised, zhang2020neuro}. The classical example is Hebbian plasticity \citep{hebb1949organization}, which describes the evolution of synapses within a single layer but does not by itself specify how networks can communicate between layers and cooperatively evolve towards a global objective.

Equilibrium Propagation (EqProp) \citep{scellier2017equilibrium} is one promising approach towards this goal. In an energy-based network whose state relaxes towards a minimum of an energy function, EqProp shows that the gradient of a global cost with respect to a synaptic parameter can be recovered from a contrastive comparison of two relaxed states, one obtained freely and one obtained while the output is nudged towards a target. The resulting update is local and has been shown to scale to deep architectures and non-trivial datasets \citep{laborieux2021scaling, scellier2024fast}. It belongs to a family of contrastive formulations that includes Contrastive Divergence
\citep{hinton2002training} and Contrastive Hebbian learning \citep{xie2003equivalence},
the latter shown to be equivalent to backpropagation in a layered network.

A central implementation challenge in EqProp is that the synaptic update depends on local quantities evaluated at two different times. A conventional realization must therefore retain the quantity reached during the free phase until the corresponding nudged value becomes available. A different solution has been demonstrated in physical contrastive learning by using two copies of the network, making the two states simultaneously available for local comparison \citep{dillavou2022demonstration}. These strategies provide the required temporal comparison through additional memory or duplicated hardware. Here we show that the same temporal information can instead be carried by the intrinsic dynamics of each synapse.

Experimental evidence \citep{wang2005coactivation, bosch2014structural, shindou2019silent} indicates that biological synapses are not fixed, static parameters, as they are often treated in machine-learning models, but instead possess rich, time-dependent internal states and multi-timescale update mechanisms. Hardware counterparts of such dynamical synapses exist as well, both CMOS-based \citep{cotteret2025distributed, atoui2025multi} and memristive \citep{wu2021volatile, sung2022simultaneous}. Together, these observations motivate us to explore whether intrinsic synaptic dynamics across different timescales can provide the temporal state storage required by EqProp.

\begin{figure}[htbp]
    \centering
    \includegraphics[width=1\linewidth]{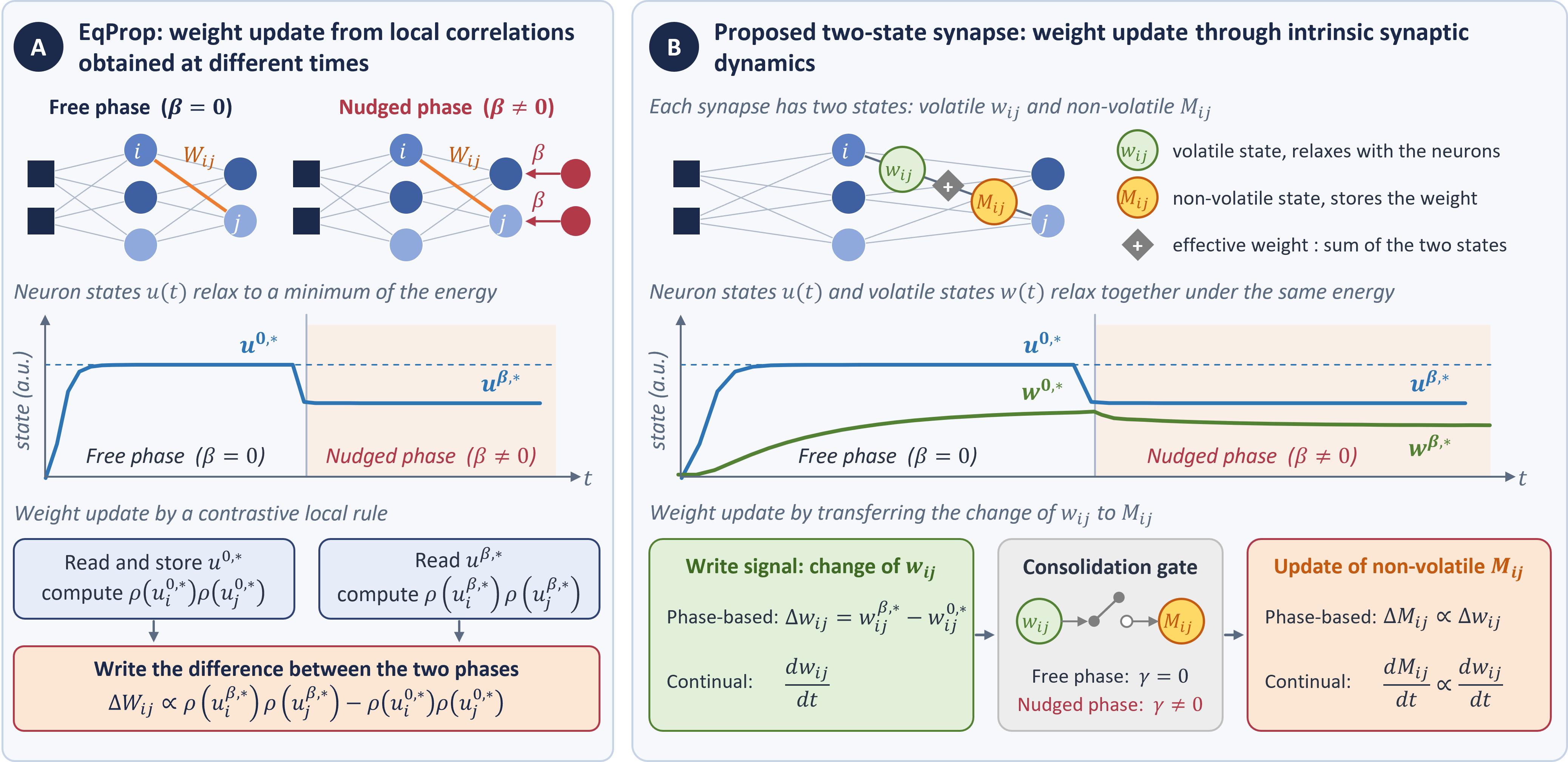}
    \caption{The internal dynamics of a two-state synapse realize the EqProp update rule. \textbf{(A)} Standard EqProp. In the free phase, only the inputs are clamped; in the nudged phase, the outputs are additionally nudged toward the target with strength $\beta$. The weight update is driven by the change of the local correlation $\rho(u_i)\rho(u_j)$ between the two phases (phase-based, shown here) or during the nudged phase (continual). \textbf{(B)} Proposed two-state synapse. Each synapse holds a volatile state $w_{ij}$ and a non-volatile state $M_{ij}$. The volatile state relaxes together with the neurons in both phases, and a consolidation gate transfers its change during the nudged phase to $M_{ij}$, either once at the end of the phase (phase-based) or continuously (continual).}
    \label{fig:dynamic_syn_illustration}
\end{figure}

We extend the network energy by introducing a volatile synaptic state $w$, in addition to the neural states, while a non-volatile state $M$ stores the learned parameter (Fig.~\ref{fig:dynamic_syn_illustration}). During the free and nudged phases, the neural states and $w$ relax together according to this energy, while $M$ is either held fixed (phase-based version) or updated slowly during the nudged phase (continual version). The volatile state $w$ relaxes toward the product of the pre- and postsynaptic activities. At equilibrium, it therefore coincides with the local term entering the EqProp weight update and acts as an intrinsic memory of that quantity. During learning, the evolution of $w$ is used to update $M$ through a global consolidation gate: transferring the total change of $w$ across the phase yields phase-based EqProp, whereas transferring its successive increments yields continual EqProp. This directly maps energy-driven synaptic dynamics onto EqProp learning and, in the continual formulation, embeds the required temporal comparison within the synaptic dynamics themselves.

\subsection{Our Main Contributions}
\emph{(1) A two-timescale synaptic substrate.} We introduce a synaptic model in which a volatile state locally tracks the pre--post correlation while a non-volatile state stores the learned parameter, and we derive the coupled neural and synaptic dynamics from a single energy function.

\emph{(2) A mapping onto Equilibrium Propagation.} We realize EqProp updates through energy-driven intrinsic synaptic dynamics. We show that the stationary volatile state provides the Hebbian term required by EqProp, and that transferring its change to the long-term state realizes both the phase-based update of \citet{scellier2017equilibrium} and the continual update of \citet{ernoult2020Ceqprop}. We further specify the conditions under which each mapping holds.

\emph{(3) A numerical proof of concept and a list of physical requirements.} We validate the proposed mapping on MNIST for both phase-based and continual updates, and use the model to derive the dynamical requirements that a volatile--non-volatile synaptic device must satisfy to implement the proposed learning mechanism.

\section{Related Work}
\label{sec:related}

\subsection{Equilibrium Propagation as a temporal synaptic update.}
\label{sec:eqprop}
\paragraph{Equilibrium Propagation.}
Energy-based learning \citep{hopfield1984neurons,hinton1984boltzmann} evolves the state of a network by descending an energy function. Equilibrium Propagation (EqProp) \citep{scellier2017equilibrium} uses this relaxation to learn from two nearby equilibria: a free equilibrium and a nudged equilibrium in which the output is weakly driven toward the target.

For reference, the network energy introduced in \citet{scellier2017equilibrium}, omitting bias terms, can be written as
\begin{equation}
    E(u) := \frac{1}{2} \sum_i u_i^2
          - \sum_{i<j}W_{ij}\,\rho(u_i)\rho(u_j),
    \label{equa:energy_sceillier_compare}
\end{equation}
where $u$ denotes the network state, $W_{ij}$ are symmetric learning parameters, and $\rho$ is the activation function. During the free phase, the network relaxes to an equilibrium $u^{0,*}$. During the nudged phase, it relaxes instead under the total energy $F=E+\beta C$, where $C$ is a cost applied to the output units and $\beta$ controls the strength of the nudging, reaching a second equilibrium $u^{\beta,*}$. The weights $W$ remain fixed during both relaxations.

The EqProp learning rule follows from comparing these two equilibria. For each synaptic weight $W_{ij}$,
\begin{equation}
    \Delta W_{ij} \propto \frac{1}{\beta}\left(
      \rho(u_i^{\beta,*})\rho(u_j^{\beta,*})
      - \rho(u_i^{0,*})\rho(u_j^{0,*})
    \right).
    \label{equa:eqprop_local_rule}
\end{equation}
The important point for a physical implementation is that the two local terms entering this update are produced at different times. The value reached during the free phase must therefore remain available until the corresponding nudged value is produced.

\paragraph{Continual Equilibrium Propagation.}
Continual EqProp \citep{ernoult2020Ceqprop} reformulates this temporal comparison by updating the weight throughout the nudged relaxation. At each step,
\begin{equation}
    \delta W_{ij}(t)\propto \frac{1}{\beta}
    \left(
      \rho(u_i^{\beta,t})\rho(u_j^{\beta,t})
      - \rho(u_i^{\beta,t-1})\rho(u_j^{\beta,t-1})
    \right).
    \label{equa:ceqprop_increment}
\end{equation}
The nudged phase starts from the free equilibrium, $u^{\beta,0}=u^{0,*}$. When the successive increments are summed, their intermediate terms cancel, leaving the difference between the initial and final states:
\begin{equation}
    \sum_{t=1}^{\infty} \delta W_{ij}(t)=\Delta W_{ij}.
    \label{equa:ceqprop_sum}
\end{equation}
Thus, continual EqProp reduces the interval over which information must be retained from an entire relaxation to a single time step. The equivalence with the phase-based rule is approximate when the updates are applied immediately, and requires the weights to change only weakly during the nudged relaxation, as discussed in \citet{ernoult2020Ceqprop}.

\paragraph{Functions required of a synapse implementing Equilibrium Propagation.}
\label{sec:requirements}

Viewed from the synapse, the phase-based and continual formulations require the same basic operations, performed over different time intervals:
\begin{enumerate}
  \item produce a local quantity from the pre- and postsynaptic activities;
  \item retain information about an earlier value of that quantity;
  \item form its change between two successive states;
  \item transfer this change to a persistent parameter during learning.
\end{enumerate}
This work investigates how these functions can be realized directly by the internal dynamics of the synaptic element.

\subsection{Synaptic dynamics in energy-based and physical learning}
\label{sec:related_physical}

Synapses with multiple internal timescales have been studied in biology \citep{wang2005coactivation, bosch2014structural, shindou2019silent}, in learning models \citep{miconi2019backpropamine,rodriguez2022shortterm}, and in neuromorphic hardware \citep{cotteret2025distributed, atoui2025multi, wu2021volatile, sung2022simultaneous}.
Fast synaptic variables have been used to retain information about past inputs, support continual learning \citep{zenke2017continual, kaplanis2018continual}, or implement eligibility traces \citep{sutton1988learning,gerstner2018eligibility, bellec2020eprop}. In our framework, the volatile state is assigned a different function: it carries the local quantity required over the temporal evolution of an EqProp update and is reset between examples. The relation with eligibility traces, which provides the closest analogy at the level of synaptic plasticity, is discussed in Section~\ref{sec:discussion_eligibility}.

A closer conceptual precedent is the work of \citet{dong1992dynamic}, where synaptic variables evolve together with neural activity and the joint dynamics decrease a Lyapunov function. Their synaptic dynamics are Hebbian-like and driven by correlations between connected neurons, providing a framework in which synaptic and neural states participate in the same energy-driven dynamics. In their work, these dynamics support self-organization. Here, we use energy-driven synaptic dynamics to carry the local temporal quantity required by EqProp and to transfer its evolution to a persistent learned parameter.

Agnostic EqProp \citep{scellier2022agnostic} provides another route for embedding learning into the physical dynamics of an energy-based system. In this approach, the synaptic parameters themselves participate in the energy minimization during the nudged phase, under a constraint that keeps them close to their previous values, enabling gradient-based learning without explicit knowledge of the system energy. In our framework, the learned parameter instead remains quasi-static during the free and nudged relaxations, while an additional volatile synaptic state participates in the energy dynamics and carries the local EqProp quantity. Its evolution is consolidated into the long-term parameter.

Latent Equilibrium \citep{haider2021latent} addresses a complementary limitation of equilibrium-based learning by defining the energy over a prospective state, allowing parameter updates before full relaxation. Our approach retains the equilibrium structure of EqProp and embeds the temporal information required by its contrastive update into the synaptic dynamics themselves.

Considerable effort has gone into building physical systems whose state settles as an energy decreases, including Ising machines \citep{laydevant2023training}, electrical resistive networks \citep{stern2022physical, Yi2022}, and mechanical metamaterials \citep{hexner2020periodic}. In most of these systems the parameter update is still computed by an external digital processor, and performing the contrastive update within the array remains open. We discuss a possible physical realization, with higher-order memristors as a promising candidate for dynamic synapses, and the necessary requirements for such a realization, in Section~\ref{sec:phy}.


\section{Two-State Synaptic Implementation of Equilibrium Propagation}

\subsection{Long-term and short-term synaptic states}

In our framework, each synapse carries two internal variables: a volatile state $w_{ij}$ and a persistent state $M_{ij}$. Table~\ref{tab:notation} summarizes the role and persistence of the different network variables.

\begin{table}[htbp]
  \centering
  \small
  \begin{tabularx}{\textwidth}{
    >{\centering\arraybackslash}p{0.1\textwidth}
    >{\centering\arraybackslash}p{0.25\textwidth}
    X
    >{\centering\arraybackslash}p{0.15\textwidth}
    >{\centering\arraybackslash}p{0.15\textwidth}
  }
    \toprule
    Symbol & Type & Role & Evolves during inference? & Persists across examples? \\
    \midrule
    $u$ & full neural state & network activity & yes & no \\
    $w$ & short-term synaptic state & local correlation trace & yes & no \\
    $M$ & long-term synaptic state & learned parameters & no & yes \\
    \bottomrule
  \end{tabularx}
  \caption{Notation of network variables. The state $u$ comprises the clamped input units together with the units that evolve.}
  \label{tab:notation}
\end{table}

The persistent state $M_{ij}$ stores the learned synaptic parameter and is retained across examples. It acts as a pure integrator and has no decay term. We therefore refer to it as long-term state: its retention is limited only by the physical element, and it accumulates over the entire training process.

The volatile state $w_{ij}$ evolves during each network relaxation, with a finite decay time and an attractor set by the instantaneous pre- and postsynaptic activities. Because it is reset between examples, we refer to it as short-term state. This distinction holds equally in the phase-based and in the continual implementation.

\subsection{The short-term timescale: relaxation of the neural and synaptic states}

\paragraph{Energy-coupled form (Variant A).}
\label{sec:volatile_dynamics}
The fixed parameter $0<g<1$ is chosen to limit the contribution of the volatile state to the effective synaptic weight $g w_{ij} + M_{ij}$.

We first consider a formulation that we call Variant A, in which the neural states and the volatile synaptic states participate in the same energy-based relaxation. The energy is defined as
\begin{equation}
E(u,w) = \frac{1}{2}\sum_i u_i^2 + \frac{g}{2}\sum_{i<j} w_{ij}^2 - \sum_{i<j}(g w_{ij}+M_{ij})\rho(u_i)\rho(u_j),
\label{equa:energy_ours}
\end{equation}
where $w_{ij}=w_{ji}$ and $M_{ij}=M_{ji}$. A fixed parameter $0<g<1$ controls the
contribution of the volatile state to the effective synaptic weight
$M_{ij}+g w_{ij}$, so that the system's  equilibrium remains dominated by the learned parameter $M_{ij}$.

The neural states $u$ and volatile synaptic variable $w$ then evolve together to minimize the network’s energy, governed by the following dynamics:

\begin{subequations}\label{equa:dynamic_set}
\begin{align}
\frac{du_i}{dt} &= -\frac{\partial E}{\partial u_i}
   = -u_i + \rho'(u_i)\sum_{j\neq i}\bigl(g\,w_{ij} + M_{ij}\bigr)\rho(u_j);
   \label{equa:dynamic_neural_A}\\
\frac{dw_{ij}}{dt} &= -\frac{1}{g}\frac{\partial E}{\partial w_{ij}}
   = -w_{ij} + \rho(u_i)\rho(u_j);
   \label{equa:dynamic_synapse}
\end{align}
\end{subequations}

Both $w$ and $u$ converge toward values corresponding to a local minimum of the energy function. As seen in Equation~\eqref{equa:dynamic_set}, the evolution of $w_{ij}$ follows a Hebbian-like update rule, driven by correlated neural activity $\rho(u_i)\rho(u_j)$, with an added decay term $w_{ij}$ to ensure stability. The non-volatile state $M$ is held fixed throughout the relaxations. The input neuron states are always clamped to the given data and do not evolve with the network dynamics.

Appendix~\ref{appendices_stability} shows that $E$ is non-increasing along \eqref{equa:dynamic_set} and bounded below, and that the fixed points are the critical points of $E$. 

\paragraph{Hardware oriented decoupled form (Variant B).}

We also consider a decoupled formulation that we call Variant B, in which the short-term state does not contribute to the effective synaptic weight. The neural dynamics are then:
\begin{equation}
    \frac{du_i}{dt} = -u_i + \rho'(u_i)\sum_{j\neq i} M_{ij}\,\rho(u_j).
\label{equa:dynamic_variantB}
\end{equation}

The neural dynamics are therefore exactly those of standard EqProp, with $M$
playing the role of the learned weight, while $w$ acts as a passive local state
that tracks the pre- and postsynaptic correlation without affecting the neural
relaxation. As a consequence, the neural subsystem retains the standard
energy-based relaxation structure of EqProp, together with its associated stability
properties \citep{scellier2017equilibrium}.

This decoupling also has two practical hardware advantages. First, only the persistent state $M$ enters the multiply--accumulate operation, avoiding the
need to combine volatile and persistent contributions in the effective weight. Second, because $w$ does not contribute to inference, its dynamics can be
disabled when learning is inactive, allowing inference to be performed using ordinary MAC operations only.

Although \(u\) and \(w\) no longer derive from a common energy in this variant, the numerical results in Section~\ref{sec:results} show that decoupling the volatile state from the neural dynamics also supports successful learning in the proposed framework.

\subsection{The long-term timescale: consolidation by phase-based and continual updates}

The update of the long-term state constitutes the learning process and follows either the phase-based formulation of \citet{scellier2017equilibrium} or the continual formulation of \citet{ernoult2020Ceqprop}. We use the same notation throughout this section: the first superscript denotes the nudging strength,
with $0$ for the free phase and $\beta$ for the nudged phase, while the second indicates when the quantity is evaluated. A star $*$ denotes the equilibrium value reached at the end of a relaxation, whereas a time index $t$ denotes a value measured during the relaxation.

\paragraph{Phase-based update.}

For the energy-coupled formulation (Variant A), we apply the EqProp result of \citet{scellier2017equilibrium} with $M$ as the trained parameter and the total energy $F=E+\beta C$, where $E$ is the augmented energy we defined in Eq.~\eqref{equa:energy_ours} and $C(u,y)$ is the cost function of the state $u$ and the target $y$. The objective is the cost at the free equilibrium, $J(M)=C\bigl(u^{0,*},y\bigr)$:
\begin{equation}
    \frac{\partial J(M)}{\partial M_{ij}} = \lim_{\beta \rightarrow 0} \frac{1}{\beta}
    \left(
      \frac{\partial F}{\partial M_{ij}}\Big|_{\beta,*}
      - \frac{\partial F}{\partial M_{ij}}\Big|_{0,*}
    \right);
\label{equa:eqprop_theorem}
\end{equation}

Since $\partial F/\partial M_{ij}=-\rho(u_i)\rho(u_j)$, the update of $M_{ij}$ takes the usual EqProp contrastive form.

The dynamic synapse provides this local correlation directly. At the end of
each relaxation, $dw_{ij}/dt=0$, so Eq.~\eqref{equa:dynamic_set} gives
\begin{equation}
\begin{aligned}
    w_{ij}^{0,*}
    &= \rho(u_i^{0,*})\rho(u_j^{0,*}),\\
    w_{ij}^{\beta,*}
    &= \rho(u_i^{\beta,*})\rho(u_j^{\beta,*}).
\end{aligned}
\label{equa:fixed_point_synapse}
\end{equation}
Thus, the local correlation entering the EqProp update is directly encoded in the stationary value of the short-term synaptic state, eliminating the need to explicitly compute and store this quantity separately.

The phase-based update can therefore be written directly in terms of the two stationary values of $w_{ij}$:
\begin{equation}
    \Delta M_{ij} = \frac{\eta_M}{\beta}\left( w_{ij}^{\beta, *} - w_{ij}^{0, *}\right),
    \label{equa:M_update_phases}
\end{equation}
where $\eta_M$ is the learning rate of the persistent state. Provided each relaxation lasts long enough for $w_{ij}$ to approach its stationary value, this update follows the gradient
of the loss, exactly in the limit $\beta\to 0$ and closely for finite $\beta$.

\paragraph{Continual update.}
\label{sec:continual_mapping}

In the continual formulation, the persistent state $M$ integrates the successive
changes of the volatile state $w$ during the nudged relaxation:
\begin{equation}
\frac{dM_{ij}}{dt}=\gamma_{0/1}\frac{\eta_M}{\beta}\frac{dw_{ij}}{dt},
    \label{equa:M_update_continous}
\end{equation}
where $\gamma_{0/1}$ is a global binary consolidation gate, equal to $0$
during the free phase and $1$ during the nudged phase.

Because the nudged relaxation starts from the free equilibrium,
$w_{ij}^{\beta,0}=w_{ij}^{0,*}$, integrating the evolution of $w_{ij}$ over
the nudged phase gives:
\begin{equation}
     \Delta M_{ij}\propto w_{ij}^{\beta, *} - w_{ij}^{0, *} = \int_{t=0}^{\infty}\frac{dw_{ij}^{\beta,t}}{dt}dt.
\end{equation}

Thus, the continual update accumulates the same net change of the volatile
state as the phase-based rule, but does so progressively during the nudged
relaxation. As in Continual EqProp \citep{ernoult2020Ceqprop}, the
correspondence with the phase-based update requires the persistent weights
to change only weakly during the relaxation.

\subsection{What the synapse stores intrinsically}
\label{sec:require_dynamic}

In conventional phase-based EqProp, the local correlation reached at the end of the free phase must remain available until the nudged equilibrium is reached. In continual EqProp, the same comparison is performed over a shorter interval:
the correlation at the current step is compared with its value at the previous
step. In both cases, a local quantity produced at an earlier time must therefore
be retained in order to form the update.

The two-state synapse uses its volatile state $w_{ij}$ to directly represent the local correlation. In the phase-based implementation, however, $w_{ij}$ evolves away from its free-equilibrium value once nudging begins. The value $w_{ij}^{0,*}$ must therefore still be retained until $w_{ij}^{\beta,*}$ is reached in order to form their difference. In the continual implementation, by contrast, the persistent state $M_{ij}$ is driven directly by $dw_{ij}/dt$ as $w_{ij}$ evolves during nudging. No value from the previous instant needs to be separately stored and read back to form the temporal
difference.

The proposed synapse therefore embeds both the local correlation and, in the continual formulation, its temporal comparison in the internal dynamics of the synaptic element. The correlation $\rho(u_i)\rho(u_j)$ is generated through the relaxation of $w_{ij}$ rather than computed and stored as a separate quantity. Table~\ref{tab:functional} summarizes which functions are performed explicitly and which are carried by the synaptic dynamics.

\begin{table}[htbp]
  \centering
  \small
  \begin{tabular}{p{3.0cm}p{3.1cm}p{3.1cm}p{3.4cm}}
    \toprule
    Function & Phase-based EqProp & Continual EqProp & Two-timescale synapse (continual update) \\
    \midrule
    Learned parameter & $W$ & $W$ & $M$ \\
    Inference weight & $W$ & $W$ & $M$ (Variant B), or $M + g\,w$ (Variant A) \\
    $\rho(u_i)\rho(u_j)$ quantity & explicit product & explicit product & relaxation of $w$ \\
    $\rho(u_i)\rho(u_j)$ retention & one full relaxation & one time step & the state of $w$ \\
    Update location & peripheral circuitry & peripheral circuitry & within the element \\
    Explicit subtraction & required & required at each step & embodied in $dw/dt$ \\
    Global control & nudging signal & nudging signal & nudging signal \& $\gamma_{0/1}$ \\
    Reset per example & $u$ & $u$ & $u$ and $w$ \\
    \bottomrule
  \end{tabular}
  \caption{Functional comparison. The table describes which operations are carried out explicitly and which are carried by an internal state. The last column refers to our continual implementation.}
  \label{tab:functional}
\end{table}


\section{Numerical Proof of Concept on MNIST}
\label{sec:results}

We train a multilayer perceptron with a single hidden layer on MNIST. These simulations serve three purposes: to verify that the two-timescale synapse preserves the learning behaviour of EqProp, to compare the energy-coupled form (Variant A) with the hardware-oriented decoupled form (Variant B), and to compare the two ways of writing the non-volatile state, phase-based and continual. Pseudocode for the latter two is given in Algorithms~\ref{pseudocode_phases} and~\ref{pseudocode_continous_update}. Numerical implementation details and training hyperparameters are illustrated in Appendix~\ref{app:simu} and ~\ref{app:hyper}.

\noindent
\begin{minipage}[t]{0.5\textwidth}
\begin{algorithm}[H]
\footnotesize
\caption{Phase-based implementation}
\begin{algorithmic}[1]
    \State \textbf{Inputs:} $X_{train}$, $M$, $\eta_M$, $\beta$, $T_{free}$, $T_{nudge}$.
    \For{epoch $n \in N$}
        \For{each image $x \in X_{train}$}
            \State Initialize neural state $\mathbf{s}$ and volatile state $\mathbf{w} \leftarrow 0$
            \For{$t = 1 \dots T_{free}$} \Comment{\textbf{Free phase}}
                \State Update $\mathbf{s}^{0,t}$
                \State Update $\mathbf{w}^{0,t}$ 
            \EndFor
            \State $\mathbf{w}^{0,*} \leftarrow \mathbf{w}^{0, T_{free}}$
            \For{$t = 1 \dots T_{nudge}$} \Comment{\textbf{Nudged phase}}
                \State Update $\mathbf{s}^{\beta,t}$ with $\beta \neq 0$
                \State Update $\mathbf{w}^{\beta,t}$
            \EndFor
            \State $\mathbf{M} \leftarrow \mathbf{M} + \dfrac{\eta_M}{\beta}\left(\mathbf{w}^{\beta,T_{nudge}} - \mathbf{w}^{0,*}\right)$
        \EndFor
    \EndFor
\end{algorithmic}
\label{pseudocode_phases}
\end{algorithm}
\end{minipage}
\hfill
\begin{minipage}[t]{0.49\textwidth}
\begin{algorithm}[H]
\footnotesize
\caption{Continual implementation}
\begin{algorithmic}[1]
    \State \textbf{Inputs:} $X_{train}$, $M$, $\eta_M$, $\beta$, $T_{free}$, $T_{nudge}$.
    \For{epoch $n \in N$}
        \For{each image $x \in X_{train}$}
            \State Initialize neural state $\mathbf{s}$ and volatile state $\mathbf{w} \leftarrow 0$
            \For{$t = 1 \dots T_{free}$} \Comment{\textbf{Free phase}}
                \State Update $\mathbf{s}^{0,t}$
                \State Update $\mathbf{w}^{0,t}$
            \EndFor
            \For{$t = 1 \dots T_{nudge}$} \Comment{\textbf{Nudged phase}}
                \State Update $\mathbf{s}^{\beta,t}$ with $\beta \neq 0$
                \State Update $\mathbf{w}^{\beta,t}$
                \State $\mathbf{M} \leftarrow \mathbf{M} + \dfrac{\eta_M}{\beta}\,(\mathbf{w}^{\beta,t}-\mathbf{w}^{\beta,t-1})$
                \State \Comment{the updated $\mathbf{M}$ is used at step $t+1$}
            \EndFor
        \EndFor
    \EndFor
\end{algorithmic}
\label{pseudocode_continous_update}
\end{algorithm}
\end{minipage}

\begin{figure}[htbp]
    \centering
    \includegraphics[width=1\linewidth]{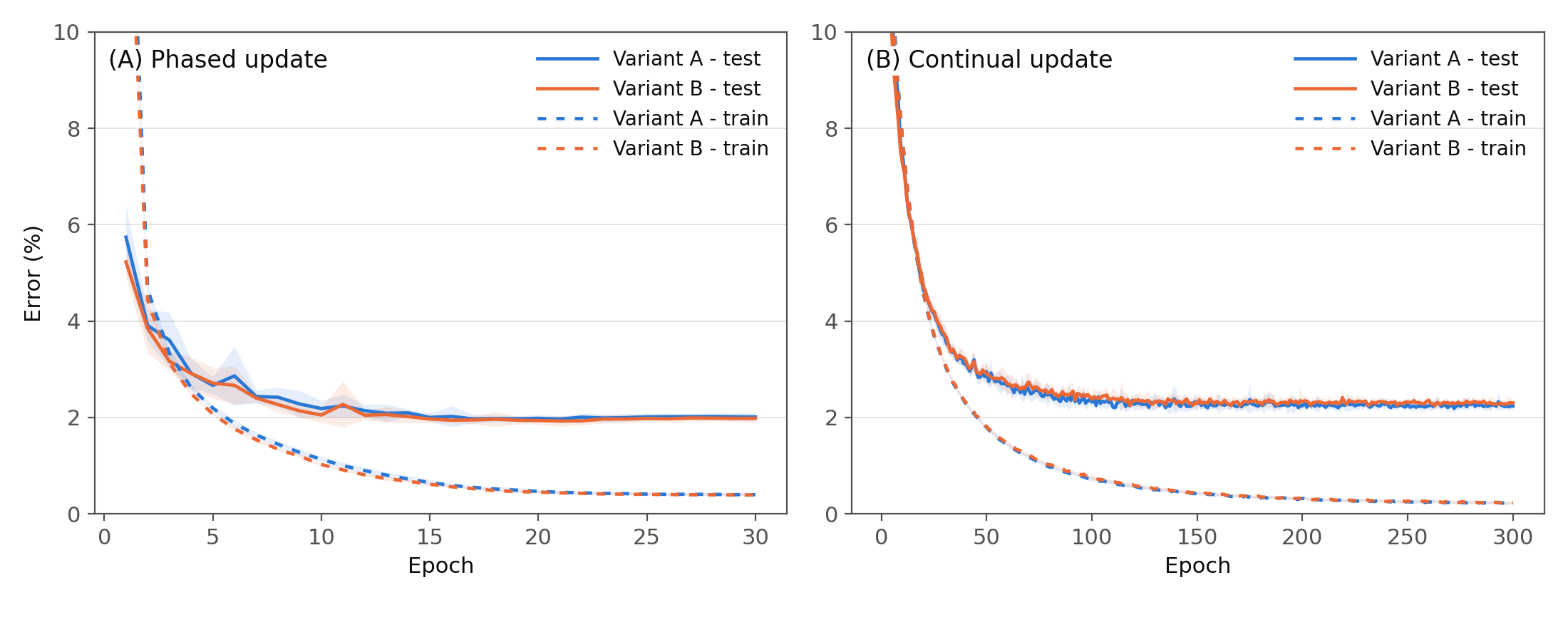}
    \caption{Learning curves of the two-state synapse variants A (blue) and B (orange) on the MNIST dataset. Depending on the persistent state $M$ update method, results are shown in subplot (A) for the phase-based update and in subplot (B) for the continual update. Solid lines show test error, while dashed lines show training error. Lines are means over 10 seeds and shaded bands are ±1 SD.}
    \label{fig:dyn_syn_learning_curves}
\end{figure}

The final training and test errors of the different methods are summarized in Table~\ref{tab:results_resume}, and Figure~\ref{fig:dyn_syn_learning_curves} shows the training and test errors over epochs for phase-based (left) and continual (right) updates. 

With phase-based updates, our method reaches a test error of 2.01\% for the energy-coupled form (Variant A) and 1.98\% for the hardware-oriented decoupled form (Variant B), compared with 2.00\% for standard phase-based EqProp. With continual updates, Variant A reaches 2.24\% and Variant B 2.30\%, compared with 2.28\% for continual EqProp. These differences lie within one standard deviation, indicating that the two-state synapse does not degrade the learning performance of EqProp on MNIST. As shown in Figure~\ref{fig:dyn_syn_learning_curves}, Variants A and B perform almost identically, which leaves more flexibility for hardware implementation.

The test error with continual updates is 0.2\% to 0.3\% higher than with phase-based updates, as also observed by \citet{ernoult2020Ceqprop}. A likely explanation
is that the learning rate must be kept small during the nudged phase to preserve the accuracy of the gradient estimate, which slows convergence and can limit the final performance. 
\begin{table}[htbp]
\centering
\begin{tabular}{@{}lclcc@{}}
\toprule
Method & Effective weight $\omega_{ij}$ & Update & Train error (\%) & Test error (\%) \\
\midrule
\multirow{2}{*}{EqProp \citep{ernoult2020Ceqprop}} & \multirow{2}{*}{$W_{ij}$}
  & Phase-based  & $\sim0.20$ & $2.00\pm0.13$ \\
& & Continual     & $\sim0.41$ & $2.28\pm0.16$ \\
\midrule
\multirow{2}{*}{Two-state synapse (Variant A)} & \multirow{2}{*}{$g\,w_{ij}+M_{ij}$}
  & Phase-based & $0.40 \pm 0.03$ & $2.01 \pm 0.09$ \\
& & Continual   & $0.22 \pm 0.03$ & $2.24 \pm 0.11$ \\
\addlinespace
\multirow{2}{*}{Two-state synapse (Variant B)} & \multirow{2}{*}{$M_{ij}$}
  & Phase-based & $0.39 \pm 0.02$ & $1.98 \pm 0.07$ \\
& & Continual   & $0.23 \pm 0.03$ & $2.30 \pm 0.09$ \\
\bottomrule
\end{tabular}
\caption{Training and test errors (\%) on MNIST for the two-state synapse and for EqProp. Our results are reported as mean $\pm$ standard deviation over 10 runs; reference EqProp results are taken from \citet{ernoult2020Ceqprop}.}
\label{tab:results_resume}
\end{table}

As a first step, these simulations indicate that our method preserves the learning capability of EqProp while using the internal synaptic dynamics to update the learned parameters.

\section{Physical implications and device requirements}
\label{sec:phy}
\subsection{Functional requirements for a two-timescale synapse}

Ideally, a single device would realize the continual implementation of the two-timescale synapse, in which the long-term state follows the short-term one throughout the nudged phase. The update would then be formed and written without leaving the array, aligning the hardware requirements for learning with those for inference.

Collecting the requirements of Section~\ref{sec:require_dynamic}, a physical element implementing the proposed substrate would have to provide:
\begin{enumerate}
\item a volatile internal state, with a finite relaxation time, driven by the activities at the two terminals;
\item a persistent, non-volatile internal state, which dominates the effective synaptic weight;
\item a transfer of the \emph{change} of the volatile state, not its value, to the non-volatile one, gated by a global signal;
\item the ability to reset the volatile state, or let it fully relax, between examples.
\end{enumerate}

\subsection{Higher-order memristive devices as candidates}
\label{sec:higher_order_realization}

Previous work has established the compatibility of EqProp with resistive physical systems, both theoretically \citep{kendall2020training, scellier2024fast} and experimentally, using resistor networks \citep{stern2022physical, dillavou2022demonstration} and first-order memristor arrays \citep{Yi2022}. These demonstrations show that the energy-based relaxation underlying EqProp can be realized directly in hardware. The remaining challenge is the temporal comparison required by the learning rule. In a single physical network, the local quantity reached during the free phase must be retained until the nudged state becomes available, which requires additional memory or peripheral circuitry. An alternative was demonstrated by \citet{dillavou2022demonstration}, where two identical resistor networks realize the free and clamped states simultaneously and local circuitry compares corresponding edges. This makes the two states simultaneously available, but requires a physical copy of the network. In the present approach, the required temporal information is instead carried by the internal dynamics of each synaptic element.

Higher-order memristors \citep{yang2018synaptic, mikheev2019ferroelectric, chaurasiya2023emerging} provide a natural candidate for such an implementation. In these devices, the conductance can depend on several coupled internal variables with distinct dynamics and retention times. Depending on the device, these variables may correspond, for example, to the geometry or ionic configuration of a conductive filament \citep{yang2012observation} and to a faster thermal state associated with Joule heating and heat dissipation \citep{kumar2022dynamical}. Such coupled electrical, structural, and thermal dynamics can generate distinct short- and long-term responses within a single element. Devices combining volatile and non-volatile internal states have already been experimentally demonstrated \citep{chang2011short, abbas2019reversible, sung2022simultaneous, Sarwat2022, weilenmann2024single}. This internal state structure provides the physical ingredients needed to represent the volatile state $w$ and the persistent learned parameter $M$ within the same synaptic element.

The mapping derived in this work specifies how these two internal states should interact. An idealized higher-order device implementing the proposed dynamics would behave as
\begin{equation}
\eqalign{
I &= G(x_1, x_2)\, V;\cr
\frac{dx_2}{dt} &= f(V) - x_2;\cr
\frac{dx_1}{dt} &\propto \gamma_{0/1}\frac{dx_2}{dt};\cr
}
\label{dyn_syn_via_high_order_memristor}
\end{equation}
where $V$ is the applied voltage, $I$ the current, and $G$ the memductance. The volatile state $x_2$ relaxes toward the voltage-dependent response $f(V)$ and plays the role of the short-term state $w$. $f(V)$ encode the local quantity as the volatage drops across the device, as illustrated in the work of \citep{kendall2020training}. The non-volatile state $x_1$ plays the role of the learned parameter $M$ and accumulates the evolution of $x_2$ only when the binary consolidation gate $\gamma_{0/1}$ is active. This gate may be implemented through an external modulation, such as a magnetic field or temperature change, that activates the long-term dynamics during learning.

Existing higher-order memristive devices already realize the coexistence of volatile and non-volatile states, but their coupling laws do not yet match those identified by the EqProp mapping. In particular, existing designs, including devices developed for eligibility-trace learning \citep{lu2022memory}, typically transfer the instantaneous value of the volatile state to the non-volatile state. The continual EqProp implementation derived here instead requires the non-volatile state to integrate the change of the volatile state, $\dot{x}_2$. In addition, the non-volatile state must dominate the effective conductance so that the learned synaptic parameter remains approximately unchanged during the free and nudged relaxations. These requirements define a concrete dynamical target for the design of a higher-order memristive synapse implementing EqProp.

\section{Discussion}
\label{sec:discuss}

\subsection{Relation to eligibility traces}
\label{sec:discussion_eligibility}

Biological synapses exhibit transient and persistent forms of plasticity operating over different timescales \citep{bosch2014structural, shindou2019silent}. A particularly
close analogy to the two-state synapse considered here is provided by eligibility traces \citep{sutton1988learning, gerstner2018eligibility}. Eligibility-trace models introduce a temporary per-synapse variable $e_{ij}$ that accumulates recent joint activity,
\begin{equation}
    \frac{de_{ij}}{dt}
    = \eta\, f_j(x_j)\,g_i(x_i)
    - \frac{e_{ij}}{\tau_e},
\end{equation}
where $x_j$ and $x_i$ denote the pre- and postsynaptic activities,
$f_j$ and $g_i$ are functions of these activities, $\eta$ sets the amplitude of the trace update, and $\tau_e$ its decay time. The trace is converted into a persistent weight change only in the presence of a third factor,
\begin{equation}
    \frac{dW_{ij}}{dt}
    = G^{\mbox{3rd}}(t)\,e_{ij},
    \label{equa:eligibility_weight_transfer}
\end{equation}
where $G^{\mbox{3rd}}(t)$ is a global modulatory signal controlling the
conversion of the eligibility trace into a weight update. Such mechanisms have experimental support \citep{bosch2014structural, shindou2019silent} and underlie learning rules such as e-prop \citep{bellec2020eprop}.

The volatile state $w_{ij}$ introduced here has a related mathematical
structure. It is a transient local variable driven by the activities of the two connected neurons and subject to relaxation, while its effect on the persistent state $M_{ij}$ is controlled by a global gate. The crucial difference lies in how the transient synaptic variable contributes to the persistent update. In a conventional eligibility-trace rule, the weight change is proportional to the trace itself. In the continual formulation derived here,
\begin{equation}
    \frac{dM_{ij}}{dt}
    \propto
    \gamma_{0/1}\frac{dw_{ij}}{dt},
    \label{equa:dynamic_synapse_weight_transfer}
\end{equation}
so that the persistent state integrates the \emph{change} of the volatile state rather than its instantaneous value. This distinction follows directly from the EqProp mapping and determines the dynamical coupling required between the volatile and persistent states.

The role of the global factor is also different. The gate $\gamma_{0/1}$
indicates whether consolidation is active during the nudged phase, but does not itself carry the sign or magnitude of the error. In addition, in the energy-coupled formulation (Variant A), the volatile state participates directly in the network energy and therefore in the neural relaxation, whereas an eligibility trace does not generally contribute to the ongoing neural computation.

Taken together, the two frameworks share a general architecture consisting of a transient local synaptic state, a persistent state, and a global signal controlling their coupling. The function assigned to the transient state is, however, specific here to EqProp: $w_{ij}$ represents the local correlation term entering the EqProp update, and its evolution during the nudged phase carries the temporal difference required by the learning rule. In the present formulation, $w_{ij}$ is reset between examples and therefore carries information over the course of a single EqProp relaxation rather than across successive inputs.

\subsection{Extensions and implementation outlook}

The continual formulation provides the most direct implementation of the proposed mechanism. Because the persistent state follows the evolution of $w_{ij}$ during the nudged relaxation, the update does not require a previous value of the local quantity to be stored separately and subsequently read back. In the phase-based formulation, by contrast, the free-equilibrium value $w_{ij}^{0,*}$ must remain available until the nudged equilibrium is reached. The two formulations therefore place different requirements on the synaptic substrate, with the continual scheme allowing the temporal comparison to be carried directly by the evolution of the volatile state.

The device-level requirements derived in Section~\ref{sec:phy} provide a basis for moving from this operating principle toward concrete implementations. A next step is to realize, or simulate at the device level, a synaptic element with the required coupling between its volatile and non-volatile states and to determine how learning is affected by realistic non-idealities, including imperfect state transfer, finite retention times, and device variability.

The separation between volatile dynamics and persistent storage also provides a natural route toward more selective consolidation. In the present model, volatile state changes occurring while the consolidation gate is active are transferred to the non-volatile state. This gate could be made conditional on additional criteria, for example by consolidating preferentially those changes that are reproduced across repeated presentations. Such a mechanism is motivated by biological observations that the transition from transient to persistent storage depends on factors such as repetition and familiarity
\citep{dudai2015consolidation, terada2022adaptive, huelin2023role, lindsey2024selective}. In a physical implementation, selective consolidation could also reduce the effect of unreliable individual writes by requiring changes to be confirmed before they are committed to the persistent state.

A further extension concerns temporal processing. In the present framework, the volatile state relaxes on the timescale of the neural dynamics and is reset between examples, so that it represents the local correlation associated with the current input. If its relaxation were made slower than the neural relaxation, $w_{ij}$ could instead integrate activity over a sequence of inputs. This would bring the framework closer to multi-timescale synaptic models developed for temporal processing, in which transient synaptic states carry information across successive inputs \citep{miconi2019backpropamine, rodriguez2022shortterm}. An EqProp-compatible formulation in this regime would require separating the neural relaxation timescale from the synaptic memory timescale and determining how the resulting history-dependent state should contribute to consolidation.


\section{Conclusion}

Building on EqProp and its continual variant, we have described a synapse with two internal states that performs, within the array, the comparison of two instants and the writing of the learning parameter. The short-term state evolves alongside the neural states to minimize the network energy in both phases. The long-term state is then driven by the change of the short-term one while a global gate is active during learning. We have confirmed on MNIST that the construction supports learning, with a decoupled configuration performing nearly as well. Higher-order memristors, which exhibit both volatile and non-volatile behavior within a single element, are a natural candidate for a future physical realization.

\section*{Conflict of Interest Statement}
The authors declare that the research was conducted in the absence of any commercial or financial relationships that could be construed as a potential conflict of interest.

\section*{Author Contributions}
The project was designed by JG and DL. The code and numerical simulations were all carried out by DL. All the authors participated to the writing of the manuscript.

\section*{Data Availability}
The programming code utilized in this study has been made publicly available online under an open-source license: \url{https://github.com/neurophysics-cnrsthales/two_state_synapses_ep}.

\section*{Funding}
This work was funded by the European Research
Council advanced grant GrenaDyn (reference: 101020684) and the public grant overseen by the French National Research Agency (ANR) as part of the ‘PEPR IA France 2030’ program (Emergences project ANR-23-PEIA-0002). The text of this
article was partially edited by a generative AI model (OpenAI ChatGPT) to improve readability.

\section*{Acknowledgments}
This project was supported with AI computing and storage resources by GENCI at IDRIS thanks to Grant 2023-AD010913993 on the supercomputer Jean Zay's V100 partition. 
\section{Appendix}

\subsection{Stability Analysis}
\label{appendices_stability}
Throughout this appendix the non-volatile state $M$ is held fixed, the couplings are symmetric, and $\rho$ is continuously differentiable with bounded range and bounded derivative. We denote the joint state by $x = (u, \tilde w)$, which collects the neural and volatile synaptic states, with the latter rescaled as $\tilde w = \sqrt{g},w$. The energy analyzed here is the joint energy used in Variant A.

\paragraph{Gradient structure.}
In the rescaled variables, the energy is rewritten as
\begin{equation}
E(x)=\frac12\sum_i u_i^{2}+\frac12\sum_{i<j}\tilde w_{ij}^{2}
-\sum_{i<j}\bigl(\sqrt{g}\,\tilde w_{ij}+M_{ij}\bigr)\rho(u_i)\rho(u_j),
\label{equa:energy_rescaled}
\end{equation}
and the joint dynamics is a gradient flow of $E$ for every $g>0$:
\begin{equation}
\frac{dx}{dt} = -\,\nabla_x E(x) .
\label{equa:gradient_flow}
\end{equation}

\paragraph{The energy decreases.}
Along Eq.\eqref{equa:gradient_flow},
\begin{equation}
\frac{dE}{dt}
 = \nabla_x E \cdot \frac{dx}{dt}
 = -\,\bigl\|\nabla_x E\bigr\|^{2}
 = -\sum_i\Bigl(\frac{du_i}{dt}\Bigr)^{2}
   -\sum_{i<j}\Bigl(\frac{d\tilde w_{ij}}{dt}\Bigr)^{2}
 \;\le\; 0 ,
\label{equa:energy_decrease}
\end{equation}
with equality if and only if $du_i/dt=0$ and $dw_{ij}/dt=0$ for all $i,j$, that is, if and only if $x$ is a critical point of $E$.

\paragraph{The energy is bounded below.}
Completing the square in each $\tilde w_{ij}$, the energy can be rewritten exactly as:
\begin{equation}
E=\frac12\|u\|^{2}
 +\frac12\sum_{i<j}\bigl(\tilde w_{ij}-\sqrt{g}\,\rho(u_i)\rho(u_j)\bigr)^{2}
 -\sum_{i<j}\Bigl[\frac g2\,\rho(u_i)^{2}\rho(u_j)^{2}+M_{ij}\,\rho(u_i)\rho(u_j)\Bigr].
\label{equa:energy_square}
\end{equation}
The first two terms are non-negative, and the last is bounded because $\rho$ is bounded
and $M$ is held fixed. Since $E$ is also non-increasing along trajectories, $E(t)$ converges.

\paragraph{Local stability.}
If the Hessian $\mathbb{H}=\nabla^{2}_{x}E(x^{*})$ at the equilibrium is positive definite, then the Jacobian at $x^{*}$ is $-\mathbb{H}$, its eigenvalues are real and strictly negative, and the equilibrium is locally asymptotically stable. The same holds in the nudged phase with $E$ replaced by
$F=E+\beta C$. A similar assumption is made in \citet{scellier2017equilibrium}. We now examine when this non-degeneracy condition holds for the two-state synapse.

\paragraph{Role of the gain $g$.}
In the rescaled variable, the Hessian has the block form
\begin{equation}
\mathbb{H}=
\begin{pmatrix} I-K & -\sqrt{g}\,J \\[2pt] -\sqrt{g}\,J^{\top} & I \end{pmatrix}.
\label{equa:hessian_blocks}
\end{equation}
This Hessian calculation requires $\rho$ to be twice continuously differentiable in a neighbourhood of the equilibrium (for the activation used in the simulations, $\rho''=0$). Writing $\rho_k=\rho(u_k^{*})$, $\rho_k'=\rho'(u_k^{*})$ and $\rho_k''=\rho''(u_k^{*})$, the matrix $K$ collects the couplings between neurons,
\begin{equation}
K_{kk}=\rho_k''\sum_l \bigl(g\,w^{*}_{kl}+M_{kl}\bigr)\rho_l,
\qquad
K_{kl}=\bigl(g\,w^{*}_{kl}+M_{kl}\bigr)\rho_k'\rho_l' \quad (k\neq l),
\end{equation}
where $g\,w^{*}_{kl}=\sqrt{g}\,\tilde w^{*}_{kl}$, and $J_{k,(ij)}=\partial[\rho(u_i)\rho(u_j)]/\partial u_k$ is the Jacobian of the correlation map. The synaptic block is the identity because each $\tilde w_{ij}$ enters the
coupling term linearly. The Schur complement therefore shows that $\mathbb{H}\succ 0$ is equivalent to $I-K-g\,JJ^{\top}\succ 0$.

We now separate the contributions of the non-volatile and volatile states by writing $K=A+g\,K_w$, where $A$ collects the terms in $M_{kl}$ and $K_w$ those in $w^{*}_{kl}$.
Combining $K_w$ with $JJ^{\top}$ into $B=K_w+JJ^{\top}$, the non-degeneracy condition becomes
\begin{equation}
I-A-gB\succ 0 .
\label{equa:nondegeneracy}
\end{equation}
Using $w^{*}_{kl}=\rho_k\rho_l$,
\begin{align}
A_{kk}&=\rho_k''\sum_{l}M_{kl}\,\rho_l, & A_{kl}&=M_{kl}\,\rho_k'\rho_l',\\
B_{kk}&=\bigl(\rho_k'^{2}+\rho_k\rho_k''\bigr)\sum_{l}\rho_l^{2}, &
B_{kl}&=2\,\rho_k\rho_k'\,\rho_l\rho_l' ,
\end{align}
with sums and off-diagonal entries restricted to connected pairs. At $g=0$ the condition reduces to $I-A\succ 0$, which is the one required in standard EqProp. Whenever it holds,
$\mathbb{H}\succ 0$ is equivalent, at the critical point considered, to
\begin{equation}
g < \frac{1}{\lambda_{\max}(S)},
\qquad
S=(I-A)^{-1/2}\,B\,(I-A)^{-1/2},
\label{equa:g_condition}
\end{equation}
any $g>0$ being admissible if $\lambda_{\max}(S)\le 0$. This criterion is local: it is checked at one equilibrium and guarantees stability only near that equilibrium. Equation~\eqref{equa:g_condition} shows that a small gain favours stability, and that the admissible range of $g$ is set by the learned parameters through $A$ and by the equilibrium state through $\rho$ and its derivatives. In practice, the condition is expected to be much milder than worst-case estimates of $\lambda_{\max}(S)$ would suggest, since $\rho'$ vanishes for saturated units and both $A$ and $B$ are sparse in a layered network.

\subsection{Implementation Details of Simulations}
\label{app:simu}

We consider a multi-layer perceptron (MLP) consisting of $N$ layers, labeled backward such that $n=0$ denotes the output layer and $n=N-1$ corresponds to the visible (input) layer. Thus, there are $N-2$ hidden layers. At time $t$, we denote the short-term (volatile) states between layer $n$ and layer $n+1$ by $\mathbf{w}_{nn+1}^{0,t}$ during the free phase, and by $w_{nn+1}^{\beta,t}$ during the nudging phase. Similarly, the neuronal activities at layer $n$, with $n\in\{0,1,\dots,N-1\}$, are denoted $\mathbf{u}_n^{0,t}$ in the free phase and $\mathbf{u}_n^{\beta,t}$ in the nudged phase. They are post-activation states and correspond to $\rho(u)$ in the main text, not to the neural potentials $u$. Additionally, we introduce long-term (non-volatile) weights between layers $n$ and $n+1$, denoted by $\mathbf{M}_{nn+1}$. These long-term weights remain fixed during the free phase. 

During our simulation, we follow the prototypical discrete-time version proposed by \cite{ernoult2019updates} to simulate the neural states dynamics. The coupled energy framework (Variant A) takes $\mathbf{M}+g\mathbf{w}$ as the effective weights. In the free phase, at each time step $t$, neural states are updated by the following equation:
\begin{equation}
\begin{aligned}
    \forall t \in [0, T_{free}] : \quad
    \begin{cases}
        \mathbf{u}_0^{0, t+1} = \sigma\left( \sigma' (\mathbf{u}_0^{0,t})(g\mathbf{w}_{01}^{0,t} + \mathbf{M}_{01}) \cdot \mathbf{u}_1^{0,t}\right), \\
        
        \mathbf{u}_n^{0, t+1} = \sigma\left(\sigma'(\mathbf{u}_n^{0, t})((g\mathbf{w}_{nn+1}^{0, t} + \mathbf{M}_{nn+1}) \cdot \mathbf{u}_{n+1}^{0, t} \right. \\
        \qquad \left. + (g(\mathbf{w}^{0,t}_{n-1n})^T+ \mathbf{M}^T_{n-1n}) \cdot \mathbf{u}_{n-1}^{0,t})\right) \forall n \in [1, N-2];
    \end{cases}
\end{aligned}
\label{equa:free_phase_neurons_rhop}
\end{equation}

During the nudged phase, hidden-layer dynamics remain unchanged, only a nudging vector is added on the output layer:
\begin{equation}
\begin{aligned}
    \forall t\in [T_{free},T_{free}+T_{nudge}] :  \quad
    \begin{cases}
        \mathbf{u}_0^{\beta, t+1} = \sigma\left(\sigma' (\mathbf{u}_0^{\beta,t})(g\mathbf{w}_{01}^{\beta, t} + \mathbf{M}_{01}) \cdot \mathbf{u}_1^{\beta, t} + \beta(\mathbf{y} - \mathbf{u}_0^{\beta, t})\right)\\
        
        \mathbf{u}_n^{\beta, t+1} = \sigma\left(\sigma' (\mathbf{u}_n^{\beta,t})((g\mathbf{w}_{nn+1}^{\beta,t} + \mathbf{M}_{nn+1}) \cdot \mathbf{u}_{n+1}^{\beta, t} \right. \\
        \qquad \left. + (g(\mathbf{w}_{n-1n}^{\beta, t})^T + \mathbf{M}^T_{n-1n}) \cdot \mathbf{u}_{n-1}^{\beta, t})\right) \forall n \in [1, N-2];
    \end{cases}
\end{aligned}
\label{equa:nudge_phase_neurons_rhop}
\end{equation}

For the hardware-oriented version (Variant B), the effective weight is simply $\mathbf{M}$. The neural dynamics are therefore obtained by setting $g=0$ in \eqref{equa:free_phase_neurons_rhop} and \eqref{equa:nudge_phase_neurons_rhop}, so that
$\mathbf{w}$ no longer appears in them. What remains is exactly the discrete-time scheme of \citet{ernoult2019updates} with the weight $W$ replaced by the long-term state $\mathbf{M}$, and we do not repeat the equations here.

The short-term state $\mathbf{w}$ follows the same update in the free and nudged phases, in both variants (A \& B):
\begin{equation}
\begin{aligned}
    \forall t \in [0, T_{free}+T_{nudge}], \; \forall n \in [0, N-2]: \\
    \mathbf{w}_{nn+1}^{t+1} = \mathbf{w}_{nn+1}^t 
    + &\Delta t \left(\mathbf{u}_n^{t+1} \cdot (\mathbf{u}_{n+1}^{t+1})^T - \mathbf{w}_{nn+1}^t\right)
\end{aligned}
\label{equa:volatile_weight_update}
\end{equation}

The phase-based update for the long-term state $\mathbf{M}$ occurs at the end of nudged phase, and it is calculated by:
\begin{equation}
    \forall n \in [0, N-2] : 
    \Delta \mathbf{M}_{nn+1} = \frac{\eta_{M}}{\beta}(\mathbf{w}_{nn+1}^{\beta,*} - \mathbf{w}_{nn+1}^{0,*})
    \label{equa:M_update_numerical_phase}
\end{equation}
where $\mathbf{w}_{nn+1}^{0,*}$ and $\mathbf{w}_{nn+1}^{\beta,*}$ are the stable states of the short-term memory at the end of the free and nudging phases, respectively.

In the continual update, $\mathbf{M}$ changes at the same rate as the short state $\mathbf{w}$ during the nudged phase, each increment differing only by the factor $\eta_M/\beta$:
\begin{equation}
    \forall n \in [0, N-2],\ \forall t \in [T_{free}, T_{free}+T_{nudge}] : \quad
    \Delta \mathbf{M}_{nn+1}^{t+1} = \frac{\eta_{M}}{\beta}
    \left(\mathbf{w}_{nn+1}^{\beta,t+1}-\mathbf{w}_{nn+1}^{\beta,t}\right).
    \label{equa:M_update_numerical_continual}
\end{equation}

\begin{figure}[htbp]
    \centering
    \includegraphics[width=1\linewidth]{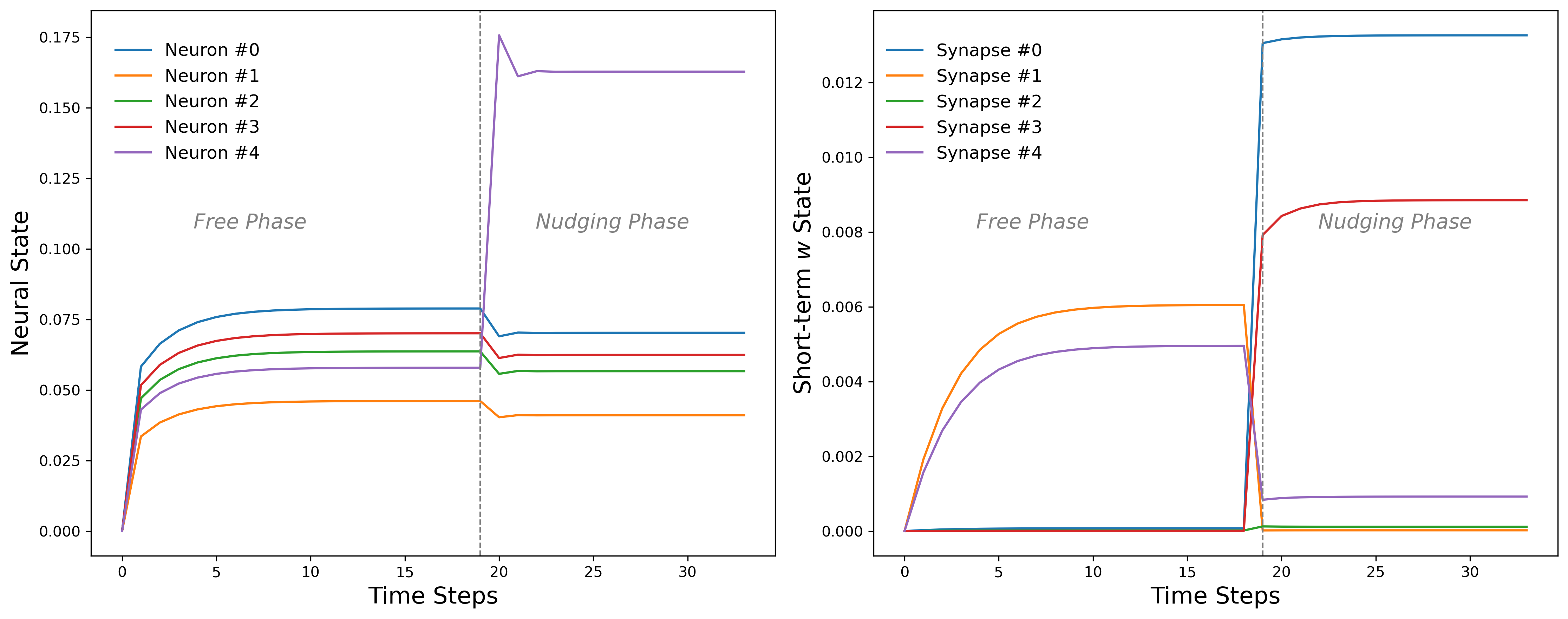}
    \caption{Evolution of neural states (left) and short-term (right) states during free phases and nudging phases.}
    \label{fig:dynamic_network_evolution}
\end{figure}

In Figure~\ref{fig:dynamic_network_evolution} we plot the dynamic evolution of neural activation and short-term memory $w$ in coupled-energy framework during the free phase and nudging. We select five random output neuron and five random $w$ at the output layer to do the illustration. The neuron nudged to the target ``1'' increase its value sharply and stabilized at a high value at the end of nudged phase. Short-term $w$ has smaller value, and has smoother evolution as it update following the derivation of energy function other than setting directly as the equilibrium point as the neural activities.

\subsection{Hyperparameters}
\label{app:hyper}
The network architecture used in the simulation results is a MLP with one hidden layer containing 512 neurons (784–512–10). Mean squared error (MSE) is used as the loss function, and gradient updates are performed using the stochastic gradient descent (SGD) optimizer without momentum. For the phase-based updates, the learning rate is decayed once per epoch with a cosine schedule, from $\eta_M$ to $0.1\,\eta_M$ over the $T$ training epochs. For the continual updates, the learning rate is kept constant, and a
per-layer norm clipping is applied to the update gradients,
$\mathrm{grad} \leftarrow \mathrm{grad}\cdot\min\bigl(1,\ 0.15/(\lVert\mathrm{grad}\rVert + 10^{-12})\bigr)$,
which prevents excessively large steps and allows a larger learning rate, speeding up
convergence. Note that the norm clipping rescales each step differently, so the continual updates no longer sum exactly to the phase-based update; the direction of each step, however, is preserved. Table~\ref{tab:hyper_SGD_dyn_syn_default} lists the hyperparameters used for the phase-based and continual updates; Variants A and B share the same values.
 

\begin{table}[htbp]
    \centering
    \begin{minipage}{0.55\textwidth}
    \resizebox{1\columnwidth}{!}{
    \begin{tabular}{ccc}
    \toprule
     Hyperparameters & Phase-based Update & Continual Update \\
    \midrule
    $T_{free}$ & 30 & 30\\
    $T_{nudge}$ & 10 & 5\\
    $\Delta t$ & 0.5 & 0.5 \\
    $\beta$ & 0.5 (random sign) & 0.5 (random sign)\\
    $g$ & 0.1 & 0.1\\
    Batch Size  & 128 & 128\\
    $\eta_{M}$ & 30-30 & 1.5-0.75\\
    Epoch & 30 & 300\\
    \toprule
    \end{tabular}
    }
    \caption{\label{tab:hyper_SGD_dyn_syn_default} Hyperparameters of Two-state dynamical synapses network. The hyperparameter $g$ is applicable only to Variant A. $\eta_M$ denotes the learning rate of the hidden-to-output weights.}
    \end{minipage}
\end{table}

\begin{figure}[htbp]
    \centering
    \includegraphics[width=0.5\linewidth]{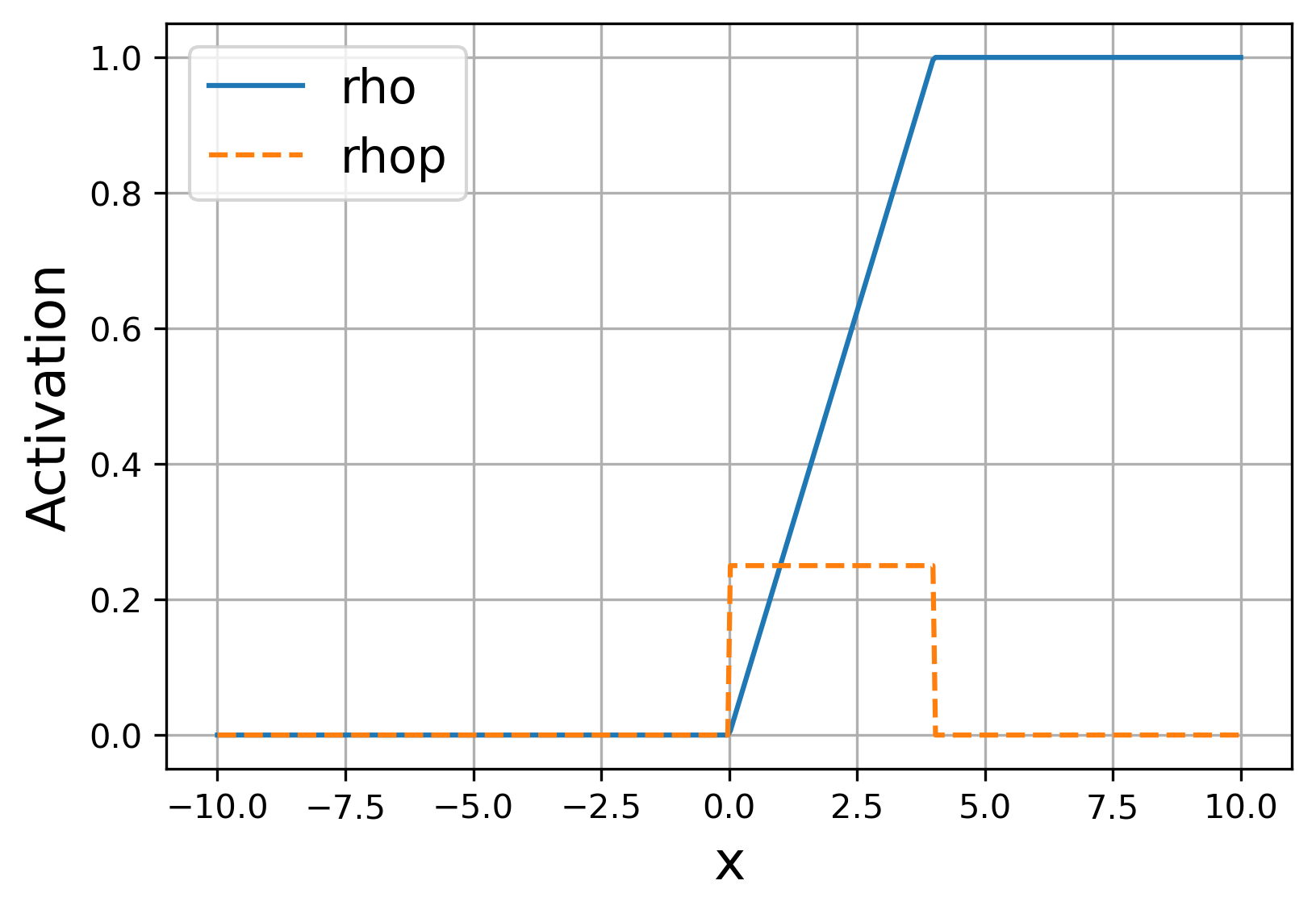}
    \caption{Activation of the proposed hard-sigmoid function with an extended linear region and its derivation.}
    \label{fig:smooth_hardsigmoid}
\end{figure}

In our dynamic synapse network, we consider a hard-sigmoid function with a wider linear regime :
\begin{verbatim}
    def rho(x):
        return torch.clamp(0.25 * x, min=0, max=1)
    def rhop(x):
        return ((x >= 0) & (x <= 4)) * 0.25
\end{verbatim}

As shown in Figure~\ref{fig:smooth_hardsigmoid}, this activation scale the input in range from 0 to 4 to the output between 0 and 1, thus can prevent the neurons achieve at the bounded value too rapidly. To prevent the derivation of activation function, one approximated way is to replace the derivation of activation function $\sigma'$ by a fixed scaling factor, noted as $s_{scale}$. By adding the scaling factor, the issue that neuron attain bounded value at the beginning thus learning fails can be effectively prevented, even with the hard-sigmoid function where non-zero input range is between 0 and 1. However, in this case, for the hidden layer the $s_{scale}$ is half of the scaling of output layer, as it receives the information from both directions.

\bibliographystyle{agsm}
\bibliography{test}

\end{document}